\documentclass[12pt]{article}
\usepackage{latexsym,amsfonts,amssymb}
\makeatletter
\@addtoreset{equation}{subsection}
\makeatother
\begin{document}
\begin{center}
{\Large \bf On Nondemolition Measurements of Photon Number:\\A
Comment on quant-ph/0612031}
\\[1.5cm]
 {\bf Vladimir S.~MASHKEVICH}\footnote {E-mail:
  Vladimir.Mashkevich100@qc.cuny.edu}
\\[1.4cm] {\it Physics Department
 \\ Queens College\\ The City University of New York\\
 65-30 Kissena Boulevard\\ Flushing, New York
 11367-1519} \\[1.4cm] \vskip 1cm

{\large \bf Abstract}
\end{center}

Experimental results stated in [1] are seminal: The authors have
realized nondemolition measurements of the photon number. As to
the interpretation of the results, it seems to be less than
convincing: The treatment of the system state and of the role of
measurement is not compatible with the conventional point of view.
We propose an adequate treatment, in which the experimental
results are a manifestation of a partial Zeno effect (a slowdown
of relaxation).

\newpage

Experimental results stated in [1] are seminal: The authors have
realized nondemolition measurements of the photon number. As to
the interpretation of the results, it seems to be less than
convincing: The treatment of the system state and of the role of
measurement is not compatible with the conventional point of view.

We quote some statements:

``A QND detection \ldots realizes an ideal projective measurement
which leaves the system in an eigenstate of the measured
observable. It can therefore be repeated many times, leading to
the same result until the system jumps into another eigenstate
under the effect of an external perturbation.''

``Our experiment realizes for the first time\ldots situation in
which the jumps of a field oscillator are revealed via QND
measurements \ldots ''

``The atoms in the QND experiment are witnessing a quantum
relaxation process whose dynamics is intrinsically not affected by
the measurement.''

These statements along with the full text imply that the field
oscillator is at any time in a pure state with a state vector
\begin{equation}\label{1}
|t\rangle:=|\mathrm{oscillator}\;t\rangle=
\sum_{n}^{0,\infty}c_{n}(t)|n\rangle
\approx\sum_{n}^{0,1}c_{n}(t)|n\rangle
\end{equation}
with
\begin{equation}\label{2}
c_{n}(t)=0\; \mathrm{or}\; 1
\end{equation}
and
\begin{equation}\label{3}
w_{n}(t)=|c_{n}(t)|^{2}\,,\quad \langle
w_{n}\rangle_{\mathrm{time}}=\lim_{T\rightarrow
\infty}=\frac{1}{T}\int\limits^{T}_{0}w_{n}(t)dt
\end{equation}
\begin{equation}\label{4}
\langle w_{n}\rangle_{\mathrm{time}}=\langle
w_{n}\rangle_{\mathrm{thermal}}\quad (\mathrm{ergodicity})
\end{equation}
\begin{equation}\label{5}
\bar{n}(t)=\langle t|\hat{n}|t\rangle\,,\quad
\hat{n}=\hat{a}^{\dag}\hat{a}
\end{equation}
\begin{equation}\label{6}
\bar{n}_{\mathrm{thermal}}:=\langle\bar{n}\rangle_{\mathrm{thermal}}\approx\langle
w_{1}\rangle_{\mathrm{thermal}}
\end{equation}

We argue that a more adequate treatment should be based on the
conventional point of view, which is this. The field oscillator is
described by a mixed state with
\begin{equation}\label{7}
\hat{\rho}(t):=\hat{\rho}_{\mathrm{oscillator}}(t)=\mathrm{Tr}_{\mathrm{thermostat}}
\{\hat{\rho}_{\mathrm{oscillator}+\mathrm{thermostat}}(t)\}
\end{equation}
In an equilibrium state,
\begin{equation}\label{8}
\hat{\rho}_{\mathrm{thermal}}=\hat{\rho}_{\mathrm{equilibrium}}=
\frac{1}{Z}\mathrm{e}^{-\hbar\omega\hat{n}/\theta}=\frac{1}{Z}
\sum_{n}^{0,\infty}\rho_{nn}|n\rangle\langle n |\,,\quad
\rho_{nn}=\mathrm{e}^{-\hbar\omega n/\theta}\,, \quad
Z=\sum_{n}^{0,\infty}\rho_{nn}
\end{equation}
and
\begin{equation}\label{9}
\mathrm{for}\;\hbar\omega\ll\theta\quad\sum_{n}^{0,\infty}\approx\sum_{n}^{0,1}
\end{equation}
In the state with a fixed $n$
\begin{equation}\label{10}
\hat{\rho}=|n\rangle\langle n|
\end{equation}

Now
\begin{equation}\label{11}
\bar{n}(t)=\mathrm{Tr}\{\hat{\rho}(t)\hat{n}\}=\sum_{n}^{0,\infty}\langle
n |\hat{\rho}(t)|n\rangle
n=\sum_{n}^{0,\infty}w_{n}(t)n=\sum_{n}^{1,\infty}w_{n}(t)n\,,\quad
w_{n}(t)=\langle n|\hat{\rho}(t)|n\rangle
\end{equation}

Let a nondemolition (nondestructive [2], ideal or of first kind
[3]) measurement be determined by a decomposition of unity
\begin{equation}\label{12}
\hat{I}=\sum_{j}\hat{P}_{j}
\end{equation}
Then the probability of a result $j$ is
\begin{equation}\label{13}
p_{j}=\mathrm{Tr}\{\hat{\rho}\hat{P}_{j}\}
\end{equation}
and
\begin{equation}\label{14}
\hat{\rho}:=\hat{\rho}(t_{\mathrm{measurement}}-0)\stackrel{\mathrm{jump}}
{\longrightarrow}\hat{\rho}(t_{\mathrm{measurement}}+0)=:\hat{\rho}'=\hat{\rho}_{j}=
\frac{\hat{P}_{j}\hat{\rho}\hat{P}_{j}}{\mathrm{Tr}\{\hat{P}_{j}\hat{\rho}\hat{P}_{j}\}}
\end{equation}
(the L$\ddot{\mathrm{u}}$ders rule [2],[3]). In our case,
\begin{equation}\label{15}
\hat{I}=\sum_{n}^{0,\infty}|n\rangle\langle n|\approx
|0\rangle\langle 0|+|1\rangle\langle 1|
\end{equation}
\begin{equation}\label{16}
\mathrm{If\, and\, only\, if}\quad \hat{\rho}=|n\rangle\langle
n|,\; \mathrm{then}\; \hat{\rho}'=\hat{\rho}\; (\mathrm{no\;
destruction})
\end{equation}

Now we use a kinetic equation for $\bar{n}(t)$:
\begin{equation}\label{17}
\frac{d\bar{n}}{dt}=B_{e}(\bar{n}+1)-B_{a}\bar{n}
\end{equation}
($e$ stands for emission, $a$ for absorption) or
\begin{equation}\label{18}
\frac{d\bar{n}}{dt}+\gamma\bar{n}=B_{e}\,,\quad\gamma=B_{a}-B_{e}
\end{equation}

The thermal value is
\begin{equation}\label{19}
\bar{n}_{\mathrm{thermal}}=\frac{B_{e}}{\gamma}=\frac{1}{B_{a}/B_{e}-1}\,,\quad
B_{e}/B_{a}=\mathrm{e}^{-\hbar\omega/\theta}\,,\quad\bar{n}_{\mathrm{thermal}}=
\frac{1}{\mathrm{e}^{\hbar\omega/\theta}-1}
\end{equation}

We obtain
\begin{equation}\label{20}
\bar{n}(t)=\bar{n}_{\mathrm{thermal}}+[\bar{n}(0)-\bar{n}_{\mathrm{thermal}}]
\mathrm{e}^{-\gamma t}
\end{equation}

Let us use the approximation (9), so that
\begin{equation}\label{21}
w_{1}(t)=\bar{n}(t)\,,\quad w_{0}(t)=1-\bar{n}(t)
\end{equation}
We have
\begin{equation}\label{22}
\mathrm{for}\;\bar{n}(0)=0\quad
w_{0}(t)=1-\bar{n}_{\mathrm{thermal}}(1-\mathrm{e}^{-\gamma
t})\,,\;\;w_{0}(0)=1
\end{equation}
\begin{equation}\label{23}
\mathrm{for}\;\bar{n}(0)=1\quad
w_{1}(t)=1-(1-\bar{n}_{\mathrm{thermal}})(1-\mathrm{e}^{-\gamma
t})\,,\;\;w_{1}(0)=1
\end{equation}
or
\begin{equation}\label{24}
w_{k}(t)=1-\langle
w_{\bar{k}}\rangle_{\mathrm{thermal}}(1-\mathrm{e}^{-\gamma
t})\,,\;\;k=0,1,\;\bar{k}=1,0,\;\;w_{k}(0)=1
\end{equation}

Consider quasicontinuous measurements as performed in [1]. Let
$\Delta t$ be the time interval between the measurements, and let
$w_{k}(0)=1$. The probability of the result $n=k\;(k=0,1)$ of the
measurement at $t=\Delta t$ is
\begin{equation}\label{25}
p_{k}(\Delta t+0)=w_{k}(\Delta t-0)=1-\langle
w_{\bar{k}}\rangle_{\mathrm{thermal}}(1-\mathrm{e}^{-\gamma\Delta
t })
\end{equation}
After $m=t/\Delta t$ measurements
\begin{equation}\label{26}
\begin{array}{l}
p_{k}(t+0)=[p_{k}(\Delta t+0)]^{m}=[1-\langle
w_{\bar{k}}\rangle_{\mathrm{thermal}}(1-\mathrm{e}^{-\gamma\Delta
t})]^{(1/\gamma\Delta t)\gamma t}\\ \qquad\qquad\, = \{[1-\langle
w_{\bar{k}}\rangle_{\mathrm{thermal}}(1-\mathrm{e}^{-x})]^{1/x}\}^{\gamma
t }\,,\quad\;\; x=\gamma\Delta t
\end{array}
\end{equation}
Now let
\begin{equation}\label{27}
x=\gamma\Delta t\ll 1\quad (\mathrm{quasicontinuity})
\end{equation}
We obtain
\begin{equation}\label{28}
[1-\langle
w_{\bar{k}}\rangle_{\mathrm{thermal}}(1-\mathrm{e}^{-x})]^{1/x}\approx
[1-\langle w_{\bar{k}}\rangle_{\mathrm{thermal}}x]^{1/x}
=\{[1-y]^{1/y}\}^{\langle
w_{\bar{k}}\rangle_{\mathrm{thermal}}}\approx\mathrm{e}^{-\langle
w_{\bar{k}} \rangle_{\mathrm{thermal}}}
\end{equation}
so that
\begin{equation}\label{29}
p_{k}(t+0)\approx\mathrm{e}^{-\langle
w_{\bar{k}}\rangle_{\mathrm{thermal}}\gamma
t}=\mathrm{e}^{-t/\tau_{k}}\,,\quad\tau=\frac{1}{\langle
w_{\bar{k}}\rangle_{\mathrm{thermal}}\gamma}\quad\mathrm{for}\;w_{k}(0)=1
\end{equation}
Thus
\begin{equation}\label{30}
\tau_{0}=\frac{1}{\bar{n}_{\mathrm{thermal}}\gamma}\,,\quad
\tau_{1}=\frac{1}{(1-\bar{n}_{\mathrm{thermal}})\gamma}
\end{equation}

Now for $T\rightarrow\infty$
\begin{equation}\label{31}
\frac{T_{n=1}}{T_{n=0}}\rightarrow\frac{\tau_{1}}{\tau_{0}}=
\frac{\bar{n}_{\mathrm{thermal}}}{1-\bar{n}_{\mathrm{thermal}}}
\end{equation}
or
\begin{equation}\label{32}
\frac{T_{n=1}}{T}\rightarrow\bar{n}_{\mathrm{thermal}}\,,\quad
\frac{T_{n=0}}{T}\rightarrow 1-\bar{n}_{\mathrm{thermal}}
\end{equation}
Thus it is the measurements that lead to the relations (32). Under
quantum jumps in the system oscillator+thermostat, $\hat{\rho}$
tends to $\hat{\rho}_{\mathrm{equilibrium}}$ (relaxation); under
those caused by the measurements of $n$, $\hat{\rho}$ undergoes
deviations from $\hat{\rho}_{\mathrm{equilibrium}}$.

Note that without measurements, i.e., in (20),(22)--(24)
\begin{equation}\label{33}
\tau=\frac{1}{\gamma}
\end{equation}
whereas with the measurements of $n$
\begin{equation}\label{34}
\tau_{k}=\frac{\tau}{\langle
w_{\bar{k}}\rangle_{\mathrm{thermal}}}>\tau
\end{equation}
This is a partial Zeno effect: a slowdown of the relaxation.

\section*{Acknowledgments}

I would like to thank Alex A. Lisyansky for support and Stefan V.
Mashkevich for helpful discussions.

\end{document}